\documentclass[amsmath,amssymb]{revtex4}

\usepackage{graphicx}
\usepackage{dcolumn}
\usepackage{bm}
\usepackage{epsfig}
\usepackage{slashed}
\usepackage{subfig}

\begin{document}

\def\0#1#2{\frac{#1}{#2}}
\def\bct{\begin{center}} \def\ect{\end{center}}
\def\beq{\begin{equation}} \def\eeq{\end{equation}}
\def\bea{\begin{eqnarray}} \def\eea{\end{eqnarray}}
\def\nnu{\nonumber}
\def\n{\noindent} \def\pl{\partial}
\def\g{\gamma}  \def\O{\Omega} \def\e{\varepsilon} \def\o{\omega}
\def\s{\sigma}  \def\b{\beta} \def\a{\alpha} \def\p{\psi} \def\r{\rho}
\def\G{\Gamma} \def\k{\kappa} \def\l{\lambda} \def\d{\delta}

\title{Quantum fluctuation in an inhomogeneous background and its influence on the phase transition in a finite volume system}
\author{Xiaogang Li$^1$}
\author{Song Shu$^1$}
\email {shus@hubu.edu.cn}
\author{Jia-Rong Li$^{2}$}

 \affiliation{1. Department of Physics and Electronic Science, Hubei
University, Wuhan, Hubei 430062, China}
 \affiliation{2. Key Laboratory of Quark and Lepton Physics (MOE) and Institute of Particle Physics, Central China Normal University, Wuhan, Hubei 430079, China}

\begin{abstract}
We have studied the grand potential and phase transitions of an inhomogeneous finite volume spherical quark system. First the finite volume effects are considered by applying the multiple reflection expansion method which is an approximation for the density of states of the momentum in the grand potential of the finite size system. Then, the density of states of momentum is further modified by the thermal fluctuations in the thermal system with the inhomogeneous field background. The modification of the density of states is calculated by the scattering phase shift from the Dirac equation of quarks in the inhomogeneous field background. By the numerical calculation we show how the phase transition is changed by varying the configuration of the inhomogeneous field background and find that the strong first order phase transition could be weakened or even changed to a crossover as a result.
\end{abstract}
\maketitle

\section{Introduction}
The physics of the phase transition in a small quantum system is very intricate, especially in a tiny droplet of quark/nuclear matter created by high energy heavy ion collisions\cite{Luo:2020pef,Stephanov:1998dy}. Quantum chromodynamics (QCD) itself and its nonperturbative nature make it very hard to solve the problem of the phase transition from the first principle\cite{Shuryak:2021vnj}. Based on the lattice QCD (LQCD) simulations and effective model studies it is known that at high temperatures and low baryon densities the transition from nuclear matter to quark matter is a smooth crossover rather than a first order phase transition. At relatively high densities the phase diagram is still unclear or not confirmed\cite{Fischer:2018sdj,Ding:2015ona,Ding:2020rtq}. The fire ball created in heavy ion collisions is very small with an estimated radius of $2\sim10\rm{fm}$. In a theoretical study one should consider the finite size effects\cite{Graef:2012sh,Klein:2017shl,Fister:2015eca,Palhares:2009tf}, and whether the system is inhomogeneous. There are many discussions about inhomogeneous phases in chiral phase transitions or deconfinement phase transitions\cite{Buballa:2014tba,Carignano:2014jla,Nickel:2009wj,Lakaschus:2020caq,Fukushima:2020cmk,Fukushima:2013rx,Kojo:2011cn}, such as chiral density wave (CDW) phases and quarkyonic phases. Additionally, the quantum effect is very hard to calculate in the inhomogeneous systems. As for the CDW phase, a plane wave chiral condensate is assumed to simplify the complicated problem of solving the quark Dirac equation in the inhomogeneous background. The quantum effect of the quark Dirac sea in an inhomogeneous background is important in studying the quark-nuclear matter phase transition. In this work we discuss the phase transition in an inhomogeneous spherical quark system which has a finite volume. The finite volume effect can be introduced in different ways. It could be added through Massachusetts Institute of Technology (MIT) bag model boundary conditions or using the multiple reflection expansion (MRE) method\cite{Xu:2020loz,Liu:2020elq,Zhao:2018nqa,Kiriyama:2002xy}. The original MRE method was derived for massless particles\cite{Balian:1970ap}; however, it has been further extended to calculate the thermodynamic quantities for the quark system with nonzero mass\cite{Madsen:1993prl,Gilson:1993prl}. It is suitable for studying a spherical system with a finite volume; thus, in this work the MRE scheme will be used as we discuss a finite spherical system. By the scheme there will be a correction to the density of states of the momentum in the momentum integration in the grand potential. It should be emphasized that previous studies of the grand potential of the thermal system often used the mean field approximation (MFA) in the homogeneous case to study the phase transition; however, in this study we argue that the quantum fluctuation over the spatially inhomogeneous background will have nontrivial influence to the phase transition. The inhomogeneous background means that in the MFA the expectation value of the field has spatial variation. As for a thermal quark system in the inhomogeneous background the usual density of states of the momentum in the standard MFA grand potential will receive a correction which can be derived by solving the quark Dirac equation to evaluate the scattering phase shift in the inhomogeneous field background. The method of evaluating the quantum fluctuations observed in the inhomogeneous field background through the scattering phase shift can be traced back to the work of Schwinger in the study of the energy of the electron's quantum fluctuations in certain nontrivial configurations of electromagnetic fields\cite{Schwinger:1954pr}. In later years Jaffe and his collaborators further completed the renormalization program and extended the method to the calculations of the quantum fluctuations in the inhomogeneous field backgrounds in electroweak systems\cite{Jaffe:1998plb,Jaffe:2002npb}. The scattering phase shift method is practical for calculating quantum fluctuations in quantum field systems with spherically inhomogeneous field backgrounds.

In this paper we will use the Friedberg-Lee (FL) model to address these problems and to illustrate the calculation scheme and its nontrivial influence to the phase transition. Previous studies of the FL model in the MFA indicate a first order phase transition at finite temperatures and densities\cite{Mao:2008prc,Gao:1992prd}. The first order phase transition is relevant to neutron star mergers and cosmological electroweak phase transitions\cite{Paschalidis:2017qmb,Ferreira:2020evu,Trodden:1998ym}. In a recent study the bubble dynamics is studied in the strong first order phase transition in the FL model\cite{Zhou:2020bzk}. In relativistic heavy ion collisions the crossover is the main feature at high temperatures and low densities. In our previous study examining the tunneling amplitude of the two vacuums in the FL model, qualitatively, we indicate that the first order phase transition may be turned into a crossover by quantum tunneling\cite{Shu:2011ij}. Several studies using chiral effective models demonstrate that quantum fluctuations may weaken the first order phase transition and strengthen the crossover feature\cite{Skokov:2010sf,Herbst:2013ail,Brandes:2021pti}. However, in these studies the quantum fluctuations are treated in a homogeneous background. In this work we will show how quantum fluctuations in an inhomogeneous background influence the phase transition in a finite volume using the FL model.

The organization of this paper is as follows: in section II the FL model is introduced and the grand potential is obtained at the MFA level. In section III it is illustrated how the grand potential is modified in a finite size quark system under an inhomogeneous background. The order parameter of the phase transition is derived accordingly. In section IV, we present the numerical results and discuss how the phase transition is changed by varying the spatial configuration of the inhomogeneous field background. The last section is the summary and outlook.

\section{The model and grand potential at MFA}
The FL model was originally introduced to describe the properties of hadrons. It incorporates different bag models, e.g. MIT and Stanford Linear Accelerator Center (SLAC) models, into one effective model of quantum field theory which is renormalizable\cite{Lee:1977prd1,Lee:1977prd2}. We start with the Lagrangian of the FL model as follows \bea {\cal
L}=\bar\psi(i\gamma_\mu\pl^\mu-g\s)\psi+\012\pl_\mu\s\pl^\mu\s-U(\s),
\eea with \bea U(\s)=\01{2!}a\s^2+\01{3!}b\s^3+\01{4!}c\s^4+B,
\eea where $\p$ is the quark field and $\s$ is the
phenomenological scalar field which can roughly describe the nonperturbative long range effect of the gluon field. $a, b, c$, and $g$ are model parameters which are fixed by fitting the general properties of nucleons. $B$ is the bag constant. In the MFA the sigma field is treated as the classical mean field or $\s(\vec r,t)\rightarrow \bar\s$. The classical potential $U(\bar\s)$ generally has two minima which correspond to two vacuums. One is the perturbative vacuum at $\bar\s=0$ which is a local minimum. The other is the physical vacuum at $\bar\s=\s_v$ which is a global minimum. If there are three valence quarks at the background of the physical vacuum they will be localized and not propagate freely in space with heavy masses. As long as the vacuum is kept in the physical vacuum, the three valence quarks in the physical ground state will be confined as a soliton which corresponds to a nucleon in a physical interpretation. Phenomenologically one can regard the nucleon as a confinement state of three valence quarks through a bag. In this model in order to produce deconfinement one has to change the vacuum. One possible way is to heat the vacuum. In the FL model one can introduce thermal grand potential through the standard method of finite temperature field theory. At finite temperature the partition function of the system is
\beq
Z=\int[id\p^{\dagger}][d\p][d\s]\exp\left[\int_0^{1/T}d\tau\int d^3
{\bf r}{\cal L}\right].
\eeq
In the MFA we can write the grand potential as
\beq
\O(\bar\s,T)=-\0{T\log Z}V=\O_{\s}+\O_{\textmd{kin}},
\eeq
where the sigma field part of the grand potential has the following form
\beq
\O_{\s}=\01V\int d^3\textbf{r}\left[\012(\nabla\bar\s)^2+U(\bar\s)\right],
\eeq
and $\O_{\textmd{kin}}$ is the kinetic part of the potential which can be written as
\beq
\O_{\textmd{kin}}=-\0TV \textmd{Trlog}\left[\01T(i\pl_0+i\g^{0}\vec\g\cdot\vec\nabla-\g^0g\bar\s)\right].
\eeq
In the homogeneous background field of sigma the grand potential can be derived as
\beq
\O_{\textmd{MFA}}(\bar\s,T)=U(\bar\s)-4N_f N_c T\int dk\r_{\textmd{hom}}(k)\log(1+e^{-E_k/T}),\label{pt_mfa}
\eeq
where $T$ is temperature, $V$ is the volume and $E_k=\sqrt{k^2+g^2\bar\s^2}$. $\r_{\textmd{hom}}$ is the density of states of momentum for the homogeneous case which is
\beq
\r_{\textmd{hom}}(k)=\01V\0{dN}{dk}=\0{k^2}{2\pi^2}.\label{density_hom}
\eeq
It should be noted that at $T=0$ the grand potential will be $\O(\bar\s,T=0)=U(\bar\s)$. The FL model at finite temperatures has been extensively studied in the literature. The general result is that when the temperature increases, the grand potential will change its configuration. As a result, the physical vacuum at $\bar\s=\s_v$ rises and becomes the local minimum, while the perturbative vacuum at $\bar\s=0$ turns into a global minimum which is the true vacuum. In the new thermal vacuum the valence quarks will propagate freely in space instead of being bounded, which means the system is deconfined. The deconfinement phase transition in the FL model is a strong first order phase transition.
\section{The finite size thermal quark system under the inhomogeneous background}
In the above discussion the thermal quark system is in the homogeneous case without boundaries. However the thermal quark/nuclear matter created in the real heavy ion collisions is always a finite size system and rather inhomogeneous. Theoretically one has to consider the corrections to the grand potential of the MFA. There are different ways to work out the corrections. In this section we will make use of the MRE method to give finite size corrections to the grand potential, while the corrections from the inhomogeneous background will be obtained through the scattering phase shift method.

First we discuss the calculation of $logZ$ in an inhomogeneous background. For a system in a spherical box with a radius $R$ the integration over the momentum in $logZ$ will be changed into a summation over the discrete momentum modes. The general form of $logZ$ by the summation over the discretized momentum modes can be written as
\beq logZ=\sum_{nlm}f({\bf k}_{nlm},T), \eeq
where $n,l,m$ are the quantum numbers of the radial, orbital angular momentum and its projection, respectively. Furthermore, we consider a spherically symmetric inhomogeneous background. The asymptotical eigenfunctions of the Hamiltonian of the system will be given by trigonometric functions with argument $kr+\delta_l(k)$. If we impose the condition that the eigenfunctions vanish at the boundary, it is found that $kR+\delta_l(k)=n_l\pi$ where $\delta_l(k)$ is the phase shift for the $l$th partial wave in the wave function at the boundary generated by the scattering of the inhomogeneous background. When the radius $R$ of the system is large enough, the states become infinitesimally close and we have\cite{Schwinger:1954pr,Jaffe:1998plb}
\beq \0{dn_l}{dk}=\0R\pi+\01\pi\0{d\delta_l(k)}{dk}. \label{density}\eeq
The density of states in momentum space for the $l$th partial wave is defined as $dn_l/dk$ divided by the volume $V$. The first term on the right hand side of the Eq. (\ref{density}) is related to the density of states of the homogeneous case by a summation over $l$ with a degeneracy factor $(2l+1)$ . If the system is homogeneous, the second term in the Eq. (\ref{density}) is zero. When the system volume goes to infinity the remaining homogeneous part will be identical as the density of states $\r_{hom}(k)$ as in Eq. (\ref{density_hom}). Now if we assume that the volume of the homogeneous system is sufficiently large at a microscopic level but finite at a macroscopic level, then by applying the MRE method the density of states will have the following form\cite{Kiriyama:2002xy,Madsen:1993prl}
\beq
\r_{\textmd{MRE}}(k)=\r_{\textmd{hom}}(k)+\Delta\r_{\textmd{MRE}}(k),
\label{rho_mre}\eeq
with
\beq
\Delta\r_{\textmd{MRE}}(k)=\0{3k}{R}f_S(\0km)+\0{6}{R^2}f_C(\0km),
\label{delta rho_mre}\eeq
where $R$ is the radius of the sphere and $m$ is the mass. On the right hand side the first and second terms are the surface and curvature contributions with the functions $f_S(k/m)$ and $f_C(k/m)$ defined as the following forms, respectively,
\beq
f_S(\0km)=-\01{8\pi}\left(1-\02{\pi}\arctan\0km\right),
\eeq
\beq
f_C(\0km)=\01{12\pi^2}\left[1-\0{3k}{2m}\left(\0{\pi}2-\arctan\0km\right)\right].
\eeq
For the homogeneous case the second term in the density of states of the Eq. (\ref{density}) has been dropped. Therefore within the framework of the MRE the grand potential of a finite size spherical system in the FL model can be written as
\beq
\O_{\textmd{MRE}}(\bar\s,T)=U(\bar\s)-4N_f N_c T\int dk\r_{\textmd{MRE}}(k)\log(1+e^{-E_k/T}). \label{omgmre}
\eeq

Next we take into account the inhomogeneous field background. The second term in Eq. (\ref{density}) should not be neglected. If we consider a spatial spherically symmetric sigma field $\bar\s(\vec r)=\bar\s(r)$, the density of states of momentum will receive a further correction derived from the second term in Eq. (\ref{density}), i.e.
\beq
\Delta\r_{\textmd{inho}}(k)=\03{4\pi^2R^3}\sum\limits_{l}(2l+1)\0{d\d_l(k)}{dk},\label{inhomrho}
\eeq
where $l$ is the quantum number of the angular momentum. In the FL model the density of states of momentum is changed due to the thermal quark scatterings on the inhomogeneous background field. $\d_l(k)$ is the scattering phase shift of quarks in the inhomogeneous background field of sigma $\bar\s(r)$. The scattering phase shift can be determined by the partial wave scattering method from the following stationary Dirac equation
\beq \left[-i\g_0\vec\g\cdot\vec\nabla+\g_0g\bar\s(r)\right]\psi=E\psi. \label{dirac} \eeq
The background field $\bar\s(r)$ can be decomposed into two parts: the homogeneous vacuum part and the inhomogeneous spatial part as the following
\beq \bar\s(r)=\s_{v0}+\tilde{\s}(r).
\eeq
By this decomposition we need to ensure that when $r\rightarrow\infty$, the background field $\bar\s(r)$ goes to its vacuum value $\s_{v0}$, and $\tilde{\s}(r)\rightarrow 0$. In this work $\tilde{\s}(r)$ is set to be a spatial configuration described using a Woods-Saxon form, which can be written as:
\beq \tilde{\s}(r)=-\0{\a}{1+e^{\b(r-r_0)}}, \label{ws} \eeq
where $\a$, $\b$ and $r_0$ are the adjustable parameters. The calculation of the scattering phase shift from Eq. (\ref{dirac}) has been discussed in our previous work \cite{Shu:2016rrr,Shu:2020hdz}, where the scattering phase shift was derived and regularized. The whole calculation of the renormalization for the phase shift is highly nontrivial. However, in the MFA limit the zero-point energy will be neglected and we only preserve the regularized phase shift in this work. In order to make the calculation self-contained, a brief derivation of the phase shift is presented in Appendix A. The final result is the regularized phase shift $\bar\d_l(k)$ which can be numerically evaluated. By substituting the regularized scattering phase shift into Eq. (\ref{inhomrho}) one can further determine the correction of the density of states due to the inhomogeneous background.

Compared to that of the homogeneous system the density of states for a finite volume inhomogeneous system will have the correction term of $\Delta\r_{\textmd{inho}}$. The final result of the density of states for the finite size inhomogeneous system is
\beq
\r(k)=\r_{\textmd{hom}}(k)+\Delta\r_{\textmd{MRE}}(k)+\Delta\r_{\textmd{inho}}(k)=\0{k^2}{2\pi^2}+\0{3k}{R}f_S(\0km)+\0{6}{R^2}f_C(\0km)+\03{4\pi^2R^3}\sum\limits_{l}(2l+1)\0{d\bar\d_l(k)}{dk},
\label{rhotot} \eeq
Then the kinetic part of the grand potential for an inhomogeneous sphere can be written as
\beq
\O_{\textmd{kin}}=-4N_f N_c T\int dk\r(k)\log(1+e^{-E_k/T}),
\eeq
where $E_k=\sqrt{k^2+g^2\bar\s^2}$. In contrast, the sigma field part of the grand potential for the inhomogeneous sphere has the following form
\beq
\O_{\s}=\0{4\pi}V\int drr^2\left[\012\left(\0{d\bar\s(r)}{dr}\right)^2+U(\bar\s(r))\right].
\eeq
According to the decomposition of the sigma field the homogeneous part can be separated from the integration and the result is
\beq
\O_{\s}=U(\bar\s)+\0{4\pi}V\int drr^2\left[\012\left(\0{d\tilde{\s}(r)}{dr}\right)^2+\tilde U(\tilde{\s}(r))\right],
\eeq
where the potential $\tilde U(\tilde{\s})$ satisfies that $\tilde U(\tilde{\s})\rightarrow 0$ when $r\rightarrow \infty$ and $\tilde{\s}\rightarrow 0$ and its form is
\beq
\tilde U(\tilde{\s}(r))=\01{2!}(a+b\s_{v0}+\012c\s_{v0}^2)\tilde\s(r)^2+\01{3!}(b+c\s_{v0})\tilde\s(r)^3+\01{4!}c\tilde\s(r)^4,
\eeq
in which $\s_{v0}$ is fixed by the absolute minimum of the $U(\bar\s)$ as
\beq
\s_{v0}=\0{3|b|}{2c}\left [1+\left (1-\0{8ac}{3b^2}\right )^{1/2}\right ].\label{sigmv0}
\eeq
The bag constant $B$ in $U(\bar\s)$ is thus fixed as
\beq
B=-(\0a{2!}\s_{v0}^2+\0b{3!}\s_{v0}^3+\0c{4!}\s_{v0}^4).\label{bagconst}
\eeq

Finally we obtain the grand potential of finite volume spherical inhomogeneous system as
\beq
\O_{\textmd{inhom}}(\bar\s,\tilde{\s},T)=U(\bar\s)+\0{4\pi}V\int drr^2\left[\012\left(\0{d\tilde{\s}(r)}{dr}\right)^2+\tilde U(\tilde{\s}(r))\right]-4N_f N_c T\int dk\r(k)\log(1+e^{-E_k/T}). \label{pt_mre_inhom}
\eeq
One should notice that the calculation of $\r(k)$ is nontrivial. In principle $\r(k)$ should vary with the quark mass $m=g\s_v$ which is temperature dependent. However by our numerical study we find that the variation of the mass in $\r(k)$ at finite temperatures for a given system size $R$ only gives a small modification to the grand potential which will be specifically discussed at the end of the next section. Therefore as an assumption in the following discussion the quark mass in $\r(k)$ could be taken as a fixed value at $m=g\s_{v0}$ and this assumption will also be addressed at the end of the next section. As the $\r(k)$ is only dependent on the spatial configuration of the sigma background field $\tilde\s(r)$ and the system size $R$, the homogeneous mean field $\bar\s$ in the grand potential could be treated as a variable. At a certain temperature the $\bar\s$ can be determined by the thermodynamic equilibrium condition as
\beq
\left. \0{\pl\O_{\textmd{inhom}}(\bar\s,\tilde{\s},T)}{\pl \bar\s}\right |_{\bar\s=\s_v}=0. \label{minimize}
\eeq
Thus, the sigma mean field value $\bar\s=\s_v$ at the global minimum of the grand potential $\O(\bar\s,\tilde{\s},T)$ is an order parameter of the phase transition in the finite volume inhomogeneous system at finite temperatures.

\section{Numerical results and discussions}
Now we will present the numerical results of the systems with the MFA and the finite-volume effect by MRE, and the system with inhomogeneous background in a finite-volume sphere respectively. The grand potential and the order parameter will be focused on in order to illustrate the phase transition. We will give the variation curves of grand potential with respect to $\bar{\s}$ at different temperatures in different conditions, and the corresponding curves of its global minimum, or the order parameter $\s_v$ with respect to $T$, will be also illustrated. The model parameters are set as $a=17.7{\rm fm}^{-2}, b=-1457.4{\rm fm}^{-1}, c=20000, g=12.16$ which are widely used in the literature\cite{Gao:1992prd,Wilets:1985prd}. Thus the quark mass $m=g\s_{v0}$ and the bag constant $B$ are determined according to (\ref{sigmv0}) and (\ref{bagconst}). The quark flavors and colors are set to be $N_f=2$ and $N_c=3$. The following numerical results are based on the assumption that the quark mass in $\r(k)$ is fixed. However at the end of this section we will give some numerical evaluations on the cases of the temperature dependent quark mass in $\r(k)$ and discuss its influence to our results.

\begin{figure}[htbp]
    \centering
    \subfloat[]{\label{omg_M+M}
    \begin{minipage}{0.45\textwidth}
        \centering
        \includegraphics[width=\textwidth]{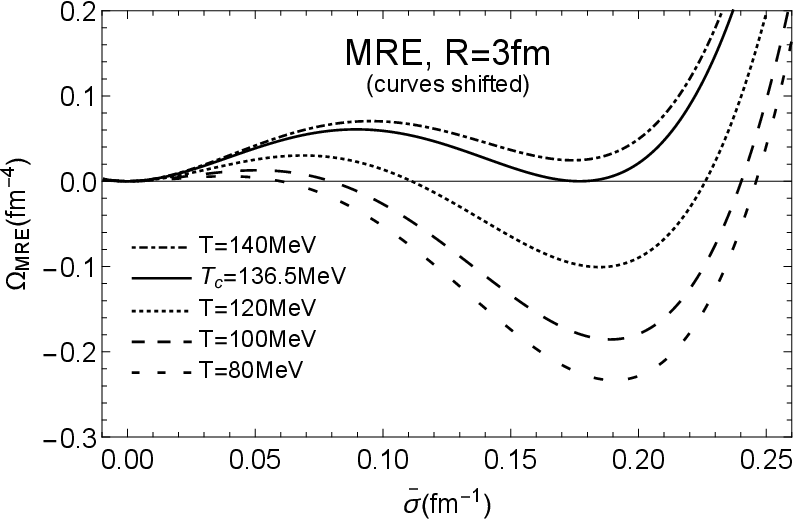}
    \end{minipage}
    }
    \qquad
    \subfloat[]{\label{sgm_M+M}
    \begin{minipage}{0.45\textwidth}
      \centering
      \includegraphics[width=\textwidth]{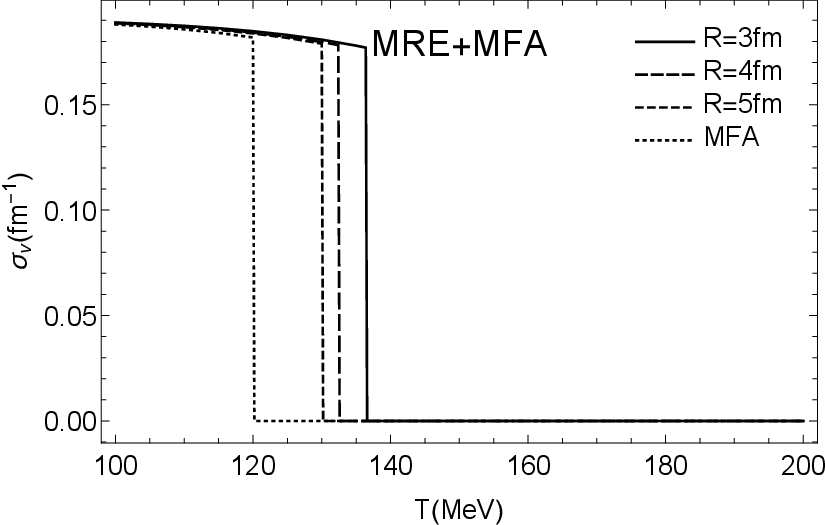}
    \end{minipage}
    }
    \captionsetup{justification=raggedright}
    \caption{(a)The variation curves of the grand potentials as functions of $\bar\s$ at $R=3\rm{fm}$ for different temperatures. (b)The variation curve of the order parameter $\s_v$ as a function of temperature $T$ for $R=3\rm{fm}, 4\rm{fm}, 5\rm{fm}$ and MFA respectively.}\label{MRE+MFA}
\end{figure}

 As the deconfinement phase transition in the FL model at MFA has been extensively studied in the literature, here we just discuss the modification to the phase transition of MFA due to the effect of MRE for the given system size $R$. From the result of the grand potential (\ref{omgmre}) with the density of states $\rho_{\rm{MRE}}$ as shown in \eqref{rho_mre}, the variation curves of the grand potentials $\O_{\rm{MRE}}(\bar{\s})$ as functions of $\bar\s $ at different temperatures at $R=3\rm{fm}$ are shown in Fig.\ref{MRE+MFA}\subref{omg_M+M}. In order to compare the potential curves at different temperatures the curves have been shifted to have the same fixed zero value of the grand potential at $\bar\s=0$. One can see there are two minima corresponding to the two vacua. At relatively low temperatures the physical vacuum is at the minimum with nonzero value of $\bar\s$. With the temperature increasing the physical vacuum rises. At a certain temperature $T_c=136.5MeV$ the two minima degenerate. The temperature $T_c$ is the critical temperature of the deconfinement phase transition at the system size of $R=3\rm{fm}$. When the temperature is further increased the physical vacuum is changed to be at the minimum with zero value of $\bar\s$ which becomes the global minimum of the grand potential. The sigma field value $\s_v$ at the physical vacuum is the order parameter, so Fig.\ref{MRE+MFA}\subref{sgm_M+M} illustrates how the order parameter varies with the temperature. The solid line corresponds to the case of the $R=3\rm{fm}$. One can see that the order parameter has a sudden jump at the critical temperature $T_c$ which represents a first order phase transition. If the system size $R$ is increased as $R=4\rm{fm}$ and $R=5\rm{fm}$, one can see that the critical temperature $T_c$ becomes lower. If it approaches the limiting case as $R\to\infty$, the MFA result which has a critical temperature $T_c=120.1MeV$ will be recovered. Thus, one can see that the description of the first order phase transition at MFA is not much changed by the finite volume effect via MRE except for a higher critical temperature at a smaller system size.

Next, we examine the numerical results of cases with the inhomogeneous background field $\tilde{\s}(r)$ for which the spatial configuration is determined by the parameters $\alpha$, $\b$ and $r_0$ as shown in \eqref{ws}. As we know at a semiclassical level there is a soliton solution in the FL model. The sigma field in the soliton solution has a spatial configuration of the Woods-Saxon type in which the values of the parameters of $\alpha$, $\b$ and $r_0$ can be fitted approximately as $\alpha\approx 0.26\rm{fm}^{-1}$, $\b\approx 8\rm{fm}^{-1}$ and $r_0\approx 1\rm{fm}$. In this study we take the configuration of the sigma field in the soliton solution as an initial configuration. The parameters $\alpha$, $\b$ and $r_0$ can vary. Our purpose is to study how the phase transition is influenced by varying the configuration of the inhomogeneous background field of the sigma. The parameter $\alpha$ describes the depth of the potential well of the sigma field. The parameters $\b$ and $r_0$ determine the spatial varying rate of the potential wall and the width of the potential well of $\tilde{\s}(r)$. It should be emphasized that in our study $\tilde{\s}(r)$ is not the meaning of the hadron but describes the spatially inhomogeneous fireball in the heavy ion collision. In the following study the vacuum value of the sigma field outside the fireball has been fixed at $\s_{v0}$. The parameter $\alpha$ will be fixed at $\alpha\approx 0.26\rm{fm}^{-1}$ which corresponds to a fixed depth of the potential well in $\tilde{\s}(r)$, while $\b$ and $r_0$ will vary by hand. The reason for this treatment is that we assume in the heavy ion collisions the difference between the vacuum values of the sigma field inside the fireball and outside the fireball will change relatively slowly during the phase transition while the size and shape of the fireball will change more remarkably.

In the calculation of the grand potential one may notice that there is a summation over angular momentum quantum number $l$ in the evaluation of the inhomogeneous part of the density of the states (\ref{rhotot}). In principle the summation index $l$ goes to infinity. However the contribution to the density of states from the scattering of the partial wave will decrease with the angular momentum quantum number $l$ increasing. As an approximation in the real numerical calculation we only do the summation over $l=0,1$ which means only the s and p partial waves have been included in the summation.

\begin{figure}[htbp]
  \centering
  \subfloat[]{\label{sgm-r_r0v_1}
  \begin{minipage}{0.45\textwidth}
  \centering
  \includegraphics[width=\textwidth]{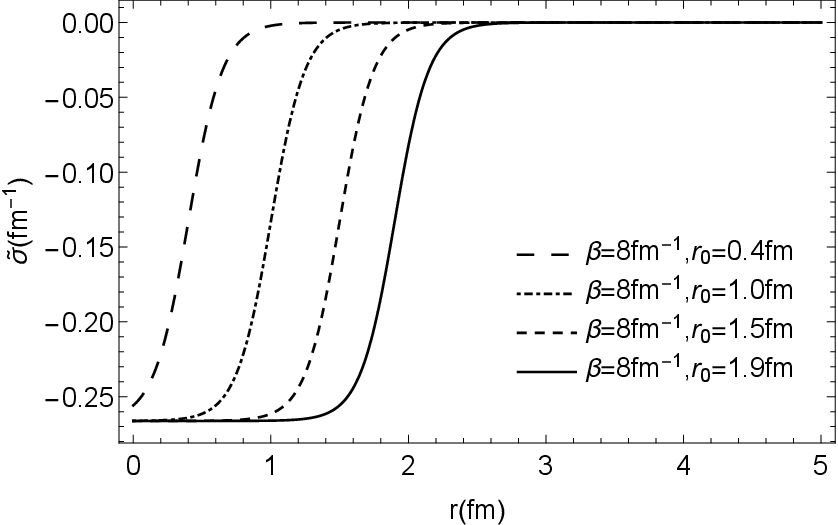}
  \end{minipage}
  }
  \quad
  \subfloat[]{\label{sgm_MI_3fm_b=8_set1}
  \begin{minipage}{0.45\textwidth}
  \centering
  \includegraphics[width=\textwidth]{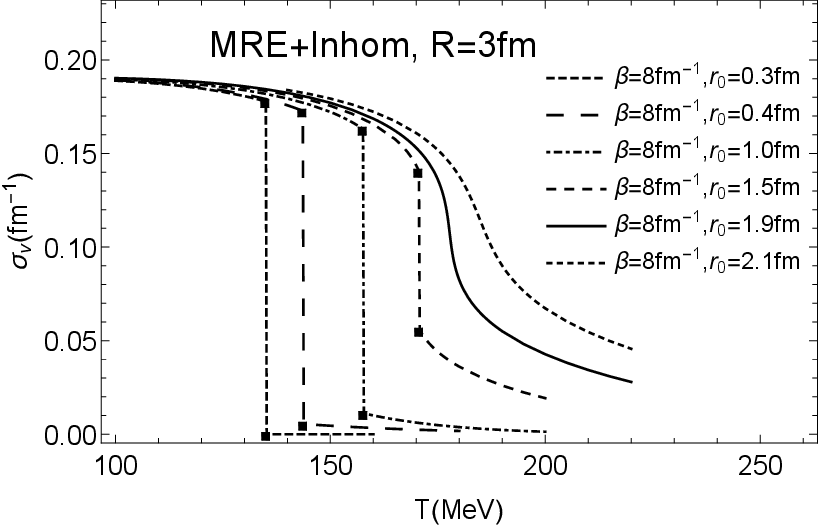}
  \end{minipage}
  }
  \captionsetup{justification=raggedright}
  \caption{(a)The spatial configurations of $\tilde{\s}(r)$ for the fixed $\b=8\rm{fm}^{-1}$ and different $r_0$. (b) The $\s_v$ as functions of temperature for different inhomogeneous backgrounds represented by different $r_0$ with fixed $\b$ and $R=3\rm{fm}$.}\label{nrm_sgmT_r0v}
\end{figure}

\begin{figure}[htbp]
  \centering
  \subfloat[]{\label{pt_MI_3fm_(8,1d5)}
  \begin{minipage}{0.45\textwidth}
  \centering
  \includegraphics[width=\textwidth]{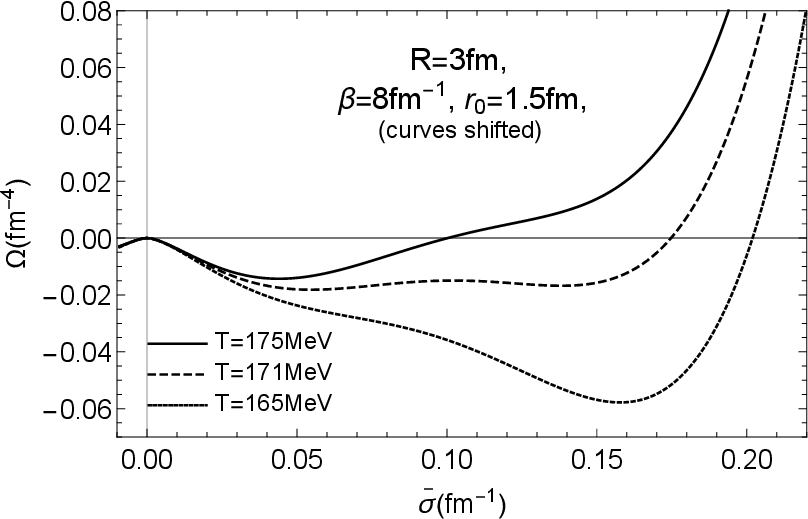}
  \end{minipage}
  }
  \quad
  \subfloat[]{\label{pt_MI_3fm_(8,2d1)}
  \begin{minipage}{0.45\textwidth}
  \centering
  \includegraphics[width=\textwidth]{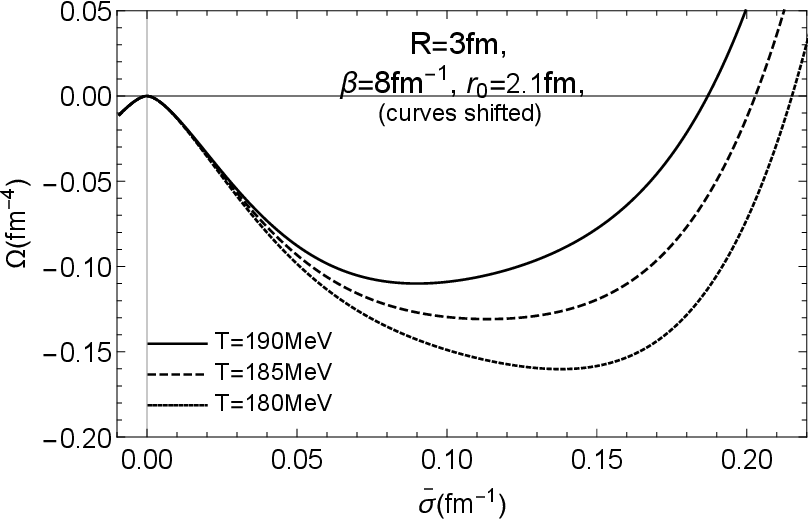}
  \end{minipage}
  }
  \captionsetup{justification=raggedright}
  \caption{(a)The grand potentials as functions of $\bar\s$ for $\b=8\rm{fm}^{-1}$ and $r_0=1.5\rm{fm}$ at $R=3\rm{fm}$ for different temperatures. (b)The grand potentials as functions of $\bar\s$ for $\b=8\rm{fm}^{-1}$ and $r_0=2.1\rm{fm}$ at $R=3\rm{fm}$ for different temperatures.}\label{pt_curv}
\end{figure}

\begin{figure}[htbp]
  \centering
  \subfloat[]{\label{ft1_(8,0d4to1d0)}
  \begin{minipage}{0.45\textwidth}
  \centering
  \includegraphics[width=\textwidth]{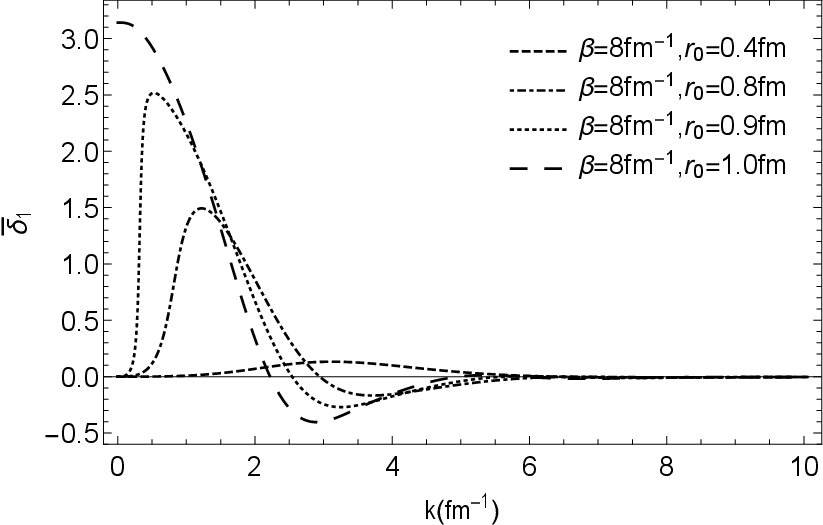}
  \end{minipage}
  }
  \quad
  \subfloat[]{\label{sgm_MI_3fm_b=8_set2}
  \begin{minipage}{0.45\textwidth}
  \centering
  \includegraphics[width=\textwidth]{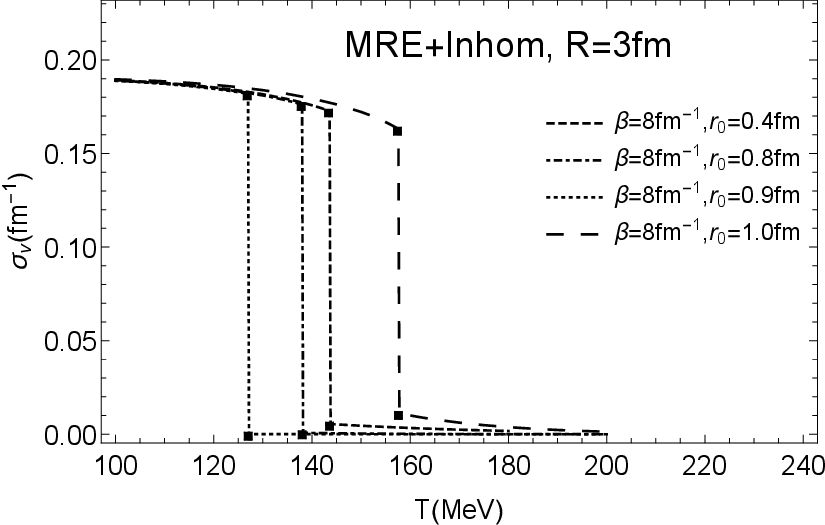}
  \end{minipage}
  }
  \captionsetup{justification=raggedright}
  \caption{The regulated p wave phase shifts as functions of momentum $k$ for fixed $\b$ and different $r_0$. (b)The $\s_v$ as functions of temperature for different inhomogeneous backgrounds represented by different $r_0$ with fixed $\b$ and $R=3\rm{fm}$.}\label{abnrm_sgmT}
\end{figure}

Now we are in a position to study how the phase transition in a finite volume is influenced by varying the spatial configuration of the background field $\tilde{\s}(r)$ through changing the parameters $\b$ and $r_0$. In the following calculation the system size is taken as $R=3\rm{fm}$. Firstly let $\b$ be fixed at $\b=8\rm{fm}^{-1}$ and vary $r_0$. From Fig. \ref{nrm_sgmT_r0v}\subref{sgm-r_r0v_1} one can see with $r_0$ increasing the width of the potential well becomes larger. By the Eq. (\ref{dirac}) the scattering phase shift can be numerically calculated in the potential background with different $r_0$, and the order parameter $\s_v$ can be determined at different temperatures through Eq. (\ref{minimize}) at certain $r_0$. In Fig.\ref{nrm_sgmT_r0v}\subref{sgm_MI_3fm_b=8_set1} the temperature dependence of the $\s_v$ for different $r_0$ has been plotted. When $r_0$ is very small ($r_0\lesssim 0.3\rm{fm}$), the spatial distribution of the inhomogeneous area is very small, and the quantum fluctuation over this tiny inhomogeneous background is also very small. As a result the phase transition is still a strong first order phase transition, and the case is very similar to that of the MRE result. When $r_0$ gets larger it is clear that quantum corrections from the inhomogeneous background become notable. In Fig.\ref{pt_curv}\subref{pt_MI_3fm_(8,1d5)} at $r_0=1.5\rm{fm}$ it is shown that besides the global minimum of the physical vacuum a new local minimum emerges near the zero value of the sigma field in the grand potential at relatively low temperatures before the phase transition. When the temperature is greater than the critical temperature this local minimum becomes the global minimum which corresponds to the physical vacuum. The corresponding variation of the order parameter $\s_v$ with the temperature could be seen in Fig.\ref{nrm_sgmT_r0v}\subref{sgm_MI_3fm_b=8_set1}. It is clear that the first order phase transition has been weakened. When $r_0$ is further increased ($r_0\gtrsim 1.9\rm{fm}$) it is found that the first order phase transition has been changed into a crossover. One can see the order parameter continuously decreasing with temperature increasing in Fig.\ref{nrm_sgmT_r0v}\subref{sgm_MI_3fm_b=8_set1}. From Fig.\ref{pt_curv}\subref{sgm_MI_3fm_b=8_set2} at $r_0=2.1\rm{fm}$ one can see that there is only one minimum in the grand potential which is corresponding to a unique vacuum in this system and with temperature increasing it continuously moves towards a small value of the sigma field which displays a crossover.
\\
In the above analysis one can see that with the spatial size of the inhomogeneous background increasing the quantum fluctuations become more and more remarkable which gradually change the first order phase transition into the crossover. Further, it is found that the critical temperature of the phase transition does not always increase with $r_0$ increasing. There is a subtle competition and balance between the contribution from the s wave scattering and p wave scattering, as follows. When $r_0\lesssim 0.3\rm{fm}$ the quantum corrections from both s wave and p wave are small. The critical temperature of the phase transition almost stays unchanged with $r_0$ increasing. When $r_0\sim0.4\rm{fm}$ there emerges an s wave bound state energy level in the energy spectrum of the Dirac Eq. (\ref{dirac}) which makes a relatively notable enhancement of the critical temperature of the phase transition as shown in Fig.\ref{nrm_sgmT_r0v}\subref{sgm_MI_3fm_b=8_set1}. At relatively small $r_0$ the quantum correction from the s wave is larger than that of the p wave. When an s wave bound orbital appears the first order phase transition is modified remarkably which means the critical temperature of the phase transition is increased and the first order phase transition is weakened. However when $r_0$ is further increased from $0.4\rm{fm}$ to $0.9\rm{fm}$ one will find that the critical temperature of the phase transition decreases and the first order phase transition strengthens as shown in Fig.\ref{abnrm_sgmT}\subref{sgm_MI_3fm_b=8_set2}. This abnormal varying trend in the above interval of varying $r_0$ is because the s wave contribution is decreasing while the p wave contribution is increasing. The contribution from the p wave counteracts that from the s wave as long as there is no bound state energy level in the p wave state. According to the Levinson index theorem the number of bound states $n_l$ with the angular momentum $l$ in our case is given by $n_l\pi=\bar\delta_l(k=0)$\cite{Ma:1984xv,Ma:2006zzc}. From Fig.\ref{abnrm_sgmT}\subref{ft1_(8,0d4to1d0)} one can see that the p wave phase shift at zero momentum remains at zero when $r_0$ changes from $0.4\rm{fm}$ to $0.9\rm{fm}$. As a result there is no bound state level in the p wave state and the p wave contribution plays the role of strengthening the first order phase transition in the above varying interval of $r_0$. However the situation will change until $r_0\gtrsim 1.0\rm{fm}$ when there emerges a bound state energy level in p wave state which can be seen from Fig.\ref{abnrm_sgmT}\subref{ft1_(8,0d4to1d0)} that $\bar\delta_1(k=0)=\pi$. There is a sign flipping of the p wave contribution such that its influence becomes greater than that of the s wave. As a result the critical temperature of the phase transition is suddenly increased dramatically and the first order phase transition is remarkably weakened which can be seen in Fig.\ref{abnrm_sgmT}\subref{sgm_MI_3fm_b=8_set2}. After that the p wave contribution increases with $r_0$ increasing and plays the dominant role in modifying the phase transition. Therefore the first order phase transition is further weakened with increasing $r_0$ and gradually turned into a crossover at $r_0\gtrsim 1.9\rm{fm}$ as shown in Fig.\ref{nrm_sgmT_r0v}\subref{sgm_MI_3fm_b=8_set1}.

\begin{figure}[htbp]
  \centering
  \subfloat[]{\label{sgm-r_bv_1}
  \begin{minipage}{0.45\textwidth}
  \centering
  \includegraphics[width=\textwidth]{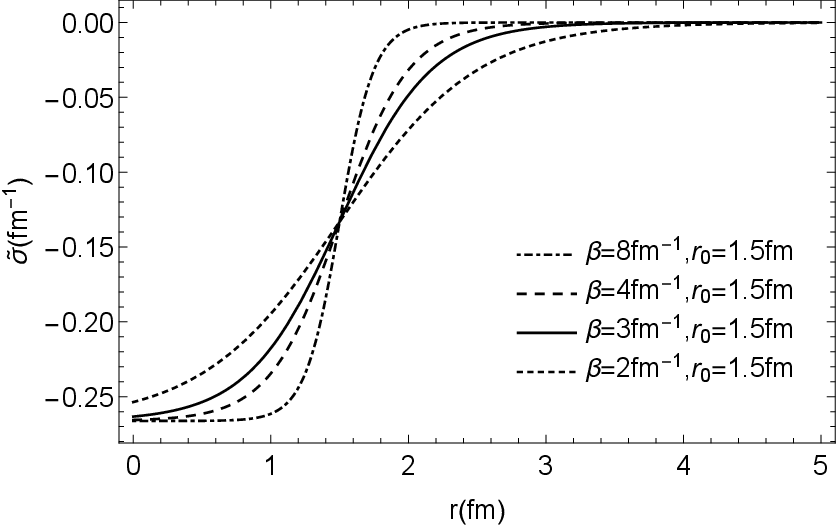}
  \end{minipage}
  }
  \quad
  \subfloat[]{\label{sgm_MI_3fm_r0=1d5}
  \begin{minipage}{0.45\textwidth}
  \centering
  \includegraphics[width=\textwidth]{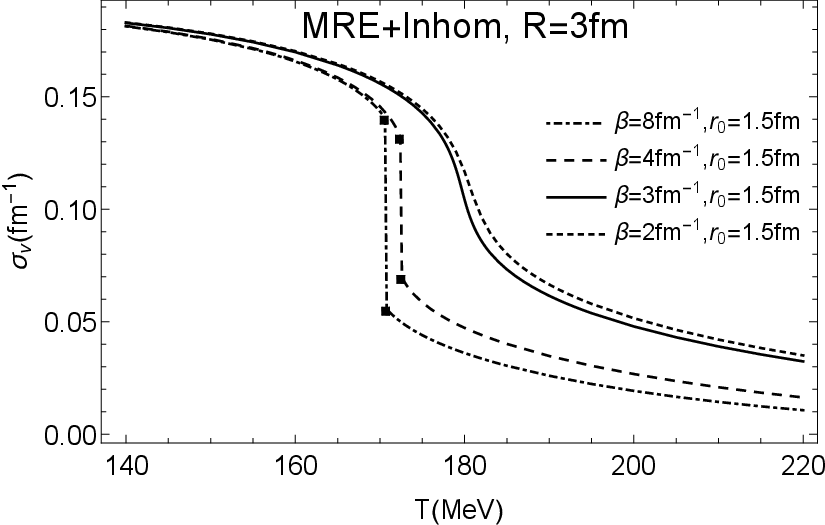}
  \end{minipage}
  }
  \captionsetup{justification=raggedright}
  \caption{(a)The spatial configurations of $\tilde{\s}(r)$ for fixed $r_0=1.5\rm{fm}$ and different $\b$. (b) The $\s_v$ as functions of temperature for different inhomogeneous backgrounds represented by different $\b$ with fixed $r_0$ and $R=3\rm{fm}$.} \label{sgmT_bv}
\end{figure}

Secondly let us fix $r_0=1.5\rm{fm}$ and let $\b$ vary. In Fig.\ref{sgmT_bv}\subref{sgm-r_bv_1} it is shown that with $\b$ decreased the variation of the sigma field over radial distance near the potential wall becomes milder. The parameter $\b$ determines the steep level of the wall of the potential well of the sigma field. It is found that with $\b$ decreased the first order phase transition is weakened and the critical temperature of the phase transition is increased as shown in Fig.\ref{sgmT_bv}\subref{sgm_MI_3fm_r0=1d5}. When $\b\gtrsim 4\rm{fm}^{-1}$ there is one bound state energy level in the s wave state and one in the p wave state. The quantum correction from the p wave scattering is more remarkable than that of the s wave. With $\b$ decreased the first order phase transition is weakened and the critical temperature of the phase transition is increased. However it should be noted that there is a nontrivial variation of the modification to the phase transition between $\b=4\rm{fm}^{-1}$ and $\b=3\rm{fm}^{-1}$. At $\b\sim3\rm{fm}^{-1}$ there emerges a second bound state level in the s wave state which can be observed from the s wave phase shift. From our numerical calculation it is found that the s wave phase shift value at zero momentum jumps from $\bar\delta_0(k=0)=\pi$ to $\bar\delta_0(k=0)=2\pi$. The correction from the s wave scattering is once again greater than that of the p wave and becomes the dominant part in the contribution to the density of the states. As a result the first order phase transition is turned into a crossover. After that the pseudo critical temperature of the crossover will slowly increase with $\b$ decreasing as long as there is no additional bound state level appearing in the s wave or p wave state.

Here we want to make some comments on the bound state energy level and how it influences the phase transition. In the inhomogeneous background field of sigma the bound state may be formed in the thermal quark system. In principle these discrete bound state energies may directly contribute additional isolated terms in the grand potential. However as these energy terms are constant energies and averaged by the space volume, they give a small constant background energy which only shifts the whole grand potential upward or downward. Therefore these isolated terms are ignored in our calculation. However the discrete bound state can also influence the grand potential by the density of states. Through the scattering mode of the continuum over the inhomogeneous background, the bound state indirectly modifies the phase transition. The phase shift of the different scattering partial wave will be changed dramatically when the bound state level emerges, so it will give remarkable modification to the density of states. As a result the grand potential will be substantially modified accordingly. Our study shows that the possible bound state in the special partial wave state in the thermal quark system will largely enhance the quantum scattering effect of the corresponding partial wave which results in weakening the first order phase transition or even turning the first order phase transition into a crossover.

We will also discuss the volume dependence of the quantum scattering effect in the inhomogeneous background. From the expression of the density of states (\ref{rhotot}) one can see that the size $R$ of the system appears in the denominator; specifically, the surface and curvature term are an order of $1/R$ and $1/R^2$ respectively, and the inhomogeneous term is order of $1/R^3$. Therefore by a qualitative analysis it can be deduced that when the system size is large enough the quantum fluctuations of the inhomogeneous background can be neglected and the MRE result will be recovered, and if the system size goes to infinity the MFA result will be recovered. In our case if the system size $R$ is greater than about $10\rm{fm}$, with $r_0$ increasing or $\b$ decreasing the first order phase transition may be weakened in some degree while the crossover will not appear. Therefore the MRE result is nearly recovered. When the system size $R$ is greater than about $20\rm{fm}$ the quantum scattering effect over the inhomogeneous background could be almost neglected and the MRE effect also becomes very small. Hence the result is very close to the case of the MFA.
\begin{figure}[htbp]
  \centering
  \subfloat[]{\label{pt_sgmv0_vs_sgmvT}
  \begin{minipage}{0.45\textwidth}
  \centering
  \includegraphics[width=\textwidth]{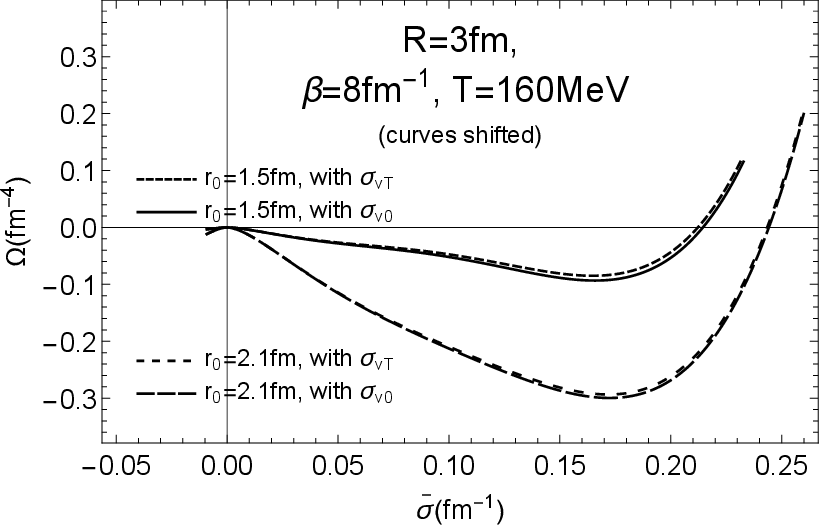}
  \end{minipage}
  }
  \quad
  \subfloat[]{\label{sgmv-T_sgmv0_vs_sgmvT}
  \begin{minipage}{0.45\textwidth}
  \centering
  \includegraphics[width=\textwidth]{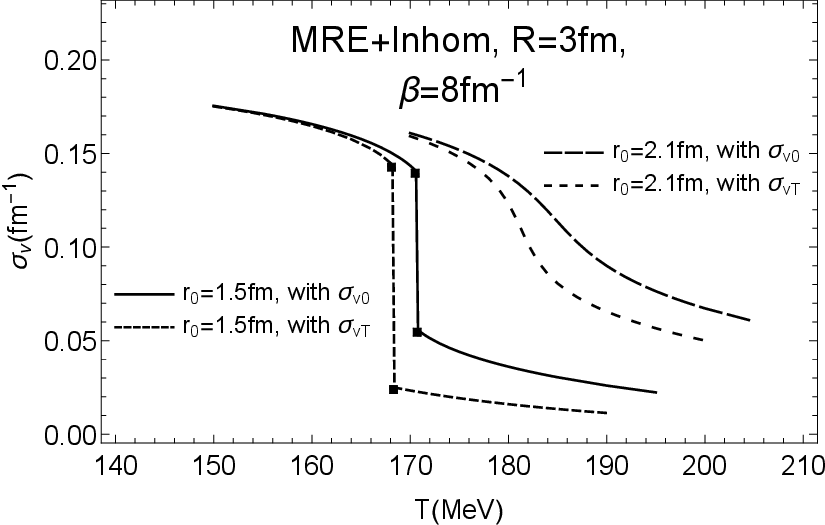}
  \end{minipage}
  }
  \captionsetup{justification=raggedright}
  \caption{(a)The grand potentials as functions of $\bar\s$ with $\s_{v0}$ vs. with $\s_{vT}$ at $\b=8{\rm fm}^{-1}$ for $r_0=1.5{\rm fm}$ and $r_0=2.1{\rm fm}$ at the temperature of $160\rm{MeV}$ and $R=3\rm{fm}$. (b)The $\s_v$ as functions of temperatures with $\s_{v0}$ vs. with $\s_{vT}$ at $\b=8{\rm fm}^{-1}$ for $r_0=1.5{\rm fm}$ and $r_0=2.1{\rm fm}$ at $R=3\rm{fm}$.}
  \label{sgmv0_vs_sgmvT}
\end{figure}

Finally we analyze the effects to our results of using the temperature dependent quark mass instead of the fixed mass in $\r(k)$. One can let the quark mass $m=g\s_v$ vary with temperature which means the $\s_v$ will not be fixed at $\s_{v0}$ but determined by the absolute minimum $\s_{vT}$ of the grand potential $\O$ at finite temperatures for the given system size $R$. This requires that the $\s_{v}$ in $\r(k)$ should be iterated when solving the Eq. (\ref{minimize}) at a given temperature. The numerical iteration process can be described, as follows. First we use $\s_{v0}$ as an initial value in $\r(k)$ to solve the Eq. (\ref{minimize}) at the given temperature. One can obtain the solution $\s_{v1}$ which is the absolute minimum of the potential $\O$. Then $\s_{v1}$ will be substituted into $\r(k)$ for the next iteration and so on. Once the difference between $\s_{vi}$ and $\s_{v,i+1}$ is very small we obtain the real stable vacuum value $\s_{vT}$ and the temperature dependent quark mass $m=g\s_{vT}$ in $\r(k)$. For an illustration of the results we have numerically evaluated two cases at the fixed $\b=8fm^{-1}$ for $r_0=1.5fm$ and $r_0=2.1fm$ respectively for the given system size $R=3\rm{fm}$. The numerical results are shown in Fig.\ref{sgmv0_vs_sgmvT}. In Fig.\ref{sgmv0_vs_sgmvT}\subref{pt_sgmv0_vs_sgmvT} the grand potentials are plotted at given temperature $T=160MeV$ with the fixed mass $m=g\s_{v0}$ and with the iterated mass $m=g\s_{vT}$ for comparisons. One can see that the grand potentials are modified very slightly due to the substitution of $\s_{v0}$ with $\s_{vT}$ in $\r(k)$. The modified degree of the grand potential will gradually increase with temperature increasing especially near the phase transition or crossover region. In Fig.\ref{sgmv0_vs_sgmvT}\subref{sgmv-T_sgmv0_vs_sgmvT} it could be seen that the order parameter determined by the absolute minimum of the grand potential has been modified accordingly. The modifications of the order parameter are more remarkable near the phase transition or crossover region than those in the regions before and after the transition or crossover. However generally speaking the quantitative modifications at finite temperatures for the finite size system are moderate and they will not qualitatively change the physical conclusions which we have made above.

\section{Summary and outlook}
In summary we have studied the grand potential and the phase transition in the thermal quark system with a finite volume under the spatially inhomogeneous background field. By varying the spatial configuration of the inhomogeneous background the scattering phase shift for the different partial wave has been numerically calculated in the corresponding inhomogeneous background at a certain small spatial volume of the system. It is found that when the bound state is formed in the thermal system the density of states of momentum will be substantially modified by the scattering phase shift of the corresponding partial wave. As a result the first order phase transition in the system with the given volume has been weakened remarkably and even turned into the crossover. However with the system volume increasing the quantum corrections in the inhomogeneous background will decrease. At sufficiently large volume of the system the quantum corrections from the inhomogeneous background can be neglected.

It should be noted that the result here may be model dependent. The different model parameters or approximation schemes may quantitatively modify the numerical results of our discussion. However the basic physical features and conclusions here will not be changed. Additionally the FL model is only a simple phenomenological model and lacks chiral symmetry. The chiral phase transition could not be studied in this model. Some outstanding questions after this study include how the scattering phase shift in the chiral models can be calculated in a spherical system with an inhomogeneous background and how the chiral phase transition will be modified by the inhomogeneous background with the possible bound states emerged. Future studies will use chiral models like the linear sigma model or the Nambu-Jona-Lasinio model to address these issues.

\appendix
\section{Derivations of the phase shift}
For the spherical symmetrical system one can assume the form of the quark field as
\beq
\psi(\vec r)=\01r\left(
\begin{array}{c}
F(r)\\
i\vec\s\cdot\vec e_r G(r)
\end{array}\right)y_{jm}^{l},
\eeq
where $y_{jm}^{l}$ is the two-component Pauli spinor harmonic and $\vec e_r$ is the spatial unit vector. $F(r)$ and $G(r)$ are the upper and lower components of the Dirac spinor. From the Dirac Eq. (\ref{dirac}) one can further derive the first order coupled differential equations of $F(r)$ and $G(r)$ as
\beq
\left(\0{d}{dr}-\0{l+1}{r}\right)F(r)+(g\bar\s(r)+E)G(r)=0,\label{u}
\eeq
\beq
\left(\0{d}{dr}+\0{l+1}{r}\right)G(r)+(g\bar\s(r)-E)F(r)=0,\label{v}
\eeq
Then the coupled equations (\ref{u}) and (\ref{v}) can be rewritten into the following decoupled second order
differential equations as \beq F''-\0{g\bar\s(r)'}{E+g\bar\s(r)}F'+\left[\0{l+1}
r\0{g\bar\s(r)'}{E+g\bar\s(r)}-\0{l(l+1)}{r^2}+(E^2-g^2\bar\s(r)^2)\right]F=0,
\label{F} \eeq \beq G''+\0{g\bar\s(r)'}{E-g\bar\s(r)}G'+\left[\0{l+1}
r\0{g\bar\s(r)'}{E-g\bar\s(r)}-\0{(l+1)(l+2)}{r^2}+(E^2-g^2\bar\s(r)^2)\right]G=0,
\label{G} \eeq where the prime denotes the differentiation with
respect to $r$. For the evaluation of the phase shift both equations will give the identical result. In the following discussion Eq.(\ref{F}) will be used for the calculation of the phase shift. When $r>>R$ the asymptotic form of Eq.(\ref{F}) is
\beq F''-\left[\0{l(l+1)}{r^2}-k^2\right]F=0, \eeq where
$k^2=E^2-g^2\s^2_{v0}$. The solutions will be spherical Hankel
functions. Meanwhile it should satisfy that $F(r)\to 0$ with $r\to 0$. There are two linearly independent solutions
\beq F^{(1)}_l(r)=e^{i\b_l(k,r)}rh^{(1)}_l(kr), \eeq \beq
F^{(2)}_l(r)=e^{-i\b^*_l(k,r)}rh^{(2)}_l(kr), \eeq where
$h^{(1)}_l(kr)$ and $h^{(2)}_l(kr)$ are the Hankel functions of
the first and second kinds and $h^{(2)}_l(kr)=h^{(1)*}_l(kr)$. The
function $\b_l(k,r)$ should satisfy $\b_l(k,r)\to 0$ as $r\to
\infty$. Then the scattering solution is \beq
F_l(r)=F^{(2)}_l(r)+e^{i2\d_l(k)}F^{(1)}_l(r), \eeq and obeys
$F_l(0)=0$, which leads to the result of the scattering phase
shift \beq \d_l(k)=-\textmd{Re}\b_l(k,0), \eeq where
$\textmd{Re}$ means the real part. It is obvious that the phase shift could be evaluated from $\b_l$. By substituting $F^{(1)}_l$
into Eq.(\ref{F}) one could obtain the equation of $\b_l$
\beq i\b''_lrh_l+2i\b'_l(h_l+rh'_l)-\b^{\prime
2}_lrh_l-\0{g\bar\s(r)'}{E+g\bar\s(r)}(i\b'_lrh_l+h_l+rh'_l)+\left[\0{l+1}{r}\0{g\bar\s(r)'}{E+g\bar\s(r)}-g^2(\bar\s(r)^2-\s^2_{v0})\right]rh_l=0.
\label{ricatti} \eeq In the fixed background field of
$\bar\s(r)$ this equation could be numerically solved to obtain the
phase shift $\d_l(k)$.

To regularize the phase shift one has to make a Born expansion which means $\b_l$ should be expanded
in powers of $g$ as \beq \b_l=g\b_{l1}+g^2\b_{l2}+\cdots.
\label{beta} \eeq Substituting the expansion (\ref{beta}) into
Eq.(\ref{ricatti}) and neglecting the higher order terms
$O(g^3)$ one can obtain a set of coupled differential equations
about $\b_{l1}$ and $\b_{l2}$ as \beq
i\b''_{l1}rh_l+(2i\b'_{l1}-\0{\bar\s(r)'}{E})(h_l+rh'_l)+(l+1)\0{\bar\s(r)'}{E}h_l=0,
\eeq \beq i\b''_{l2}rh_l-\b^{\prime
2}_{l1}rh_l-\0{i\bar\s(r)'}{E}\b'_{l1}rh_l+
(2i\b'_{l2}+\0{\bar\s(r)'\bar\s(r)}{E^2})(h_l+rh'_l)-\left[\0{(l+1)}r\0{\bar\s(r)'\bar\s(r)}{E^2}+(\bar\s(r)^2-\s^2_{v0})\right]rh_l=0.
\eeq These equations could be numerically solved to obtain the
first and second Born approximations of the phase shifts namely
$\d^{(1)}_l$ and $\d^{(2)}_l$ as \beq
\d^{(1)}_l=-g\textmd{Re}\b_{l1}(k,r=0), \ \ \
\d^{(2)}_l=-g^2\textmd{Re}\b_{l2}(k,r=0). \eeq
Finally the regularized phase shift can be defined as
\beq \bar\d_l(k)\equiv\d_l(k)-\d_l^{(1)}(k)-\d_l^{(2)}(k), \label{deltabar} \eeq
which can be numerically calculated.

\end{document}